\documentclass[aps,prl,twocolumn,showpacs,amsmath,amssymb,preprintnumbers,superscriptaddress,10pt]{revtex4-1}

\usepackage{amsmath,amssymb}
\usepackage{bm}% bold math
\usepackage[latin1]{inputenc}
\usepackage{graphicx}% include figure files
\usepackage{SIunits}
\usepackage{hyphenat}
\usepackage{hyperref}

\begin{document}

\title{Laser damage helps the eavesdropper in quantum cryptography}

\author{Audun Nystad Bugge}
\affiliation{Department of Electronics and Telecommunications, Norwegian University of Science and Technology, NO-7491 Trondheim, Norway}

\author{Sebastien Sauge}
\affiliation{School of Information and Communication Technology, Royal Institute of Technology (KTH), Electrum 229, SE-16440 Kista, Sweden}

\author{Aina Mardhiyah M.~Ghazali}
\affiliation{Department of Science in Engineering, Faculty of Engineering, International Islamic University Malaysia, P.O.\ Box 10, 50728 Kuala Lumpur, Malaysia}

\author{Johannes Skaar}
\author{Lars Lydersen}
\affiliation{Department of Electronics and Telecommunications, Norwegian University of Science and Technology, NO-7491 Trondheim, Norway}

\author{Vadim Makarov}
\email{makarov@vad1.com}
\affiliation{Institute for Quantum Computing, University of Waterloo, Waterloo, ON, N2L~3G1 Canada}

\date{February 21, 2014}

\begin{abstract}
We propose a class of attacks on quantum key distribution (QKD) systems where an eavesdropper actively engineers new loopholes by using damaging laser illumination to permanently change properties of system components. This can turn a perfect QKD system into a completely insecure system. A proof-of-principle experiment performed on an avalanche photodiode-based detector shows that laser damage can be used to create loopholes. After $\sim 1\,\watt$ illumination, the detectors' dark count rate reduces 2--5 times, permanently improving single-photon counting performance. After $\sim 1.5\,\watt$, the detectors switch permanently into the linear photodetection mode and become completely insecure for QKD applications.
\end{abstract}
% PACS, the Physics and Astronomy Classification Scheme

\maketitle

\begin{figure*}
\includegraphics[width=127mm]{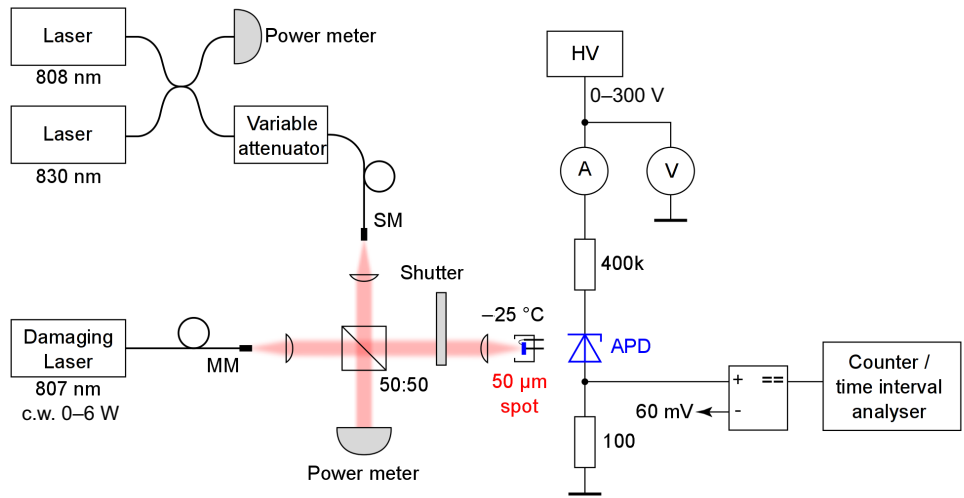}
\caption{\label{fig:setup} Experimental setup for APD damage and characterization. See text for details.}
\end{figure*}

Quantum key distribution (QKD) enables two remote parties to grow a secret key \cite{bennett1984}. The security relies on the laws of physics, provided the components and the system behave according to the models in the security proof \cite{mayers1996,lo1999,shor2000}. Practical implementations contain imperfections, however, which may enable so-called quantum hacking attacks \cite{makarov2006,zhao2008,lydersen2010a,gerhardt2011}. Work is now in progress to restore security, by modifying the implementations to avoid large loopholes \cite{lydersen2011a,yuan2011,lo2012,braunstein2012,rubenok2012}, and generalizing the security proofs \cite{mayers1996,gottesman2004,koashi2009,maroy2010,tomamichel2012} to take the remaining, unavoidable imperfections into account \footnote{It is still necessary for the manufacturer to show that the actual system is within the more general models of these security proofs. This is typically obtained by characterizing the system in a controlled environment. Note also however that the manufacturers could in turn rely on the specifications of the components, thereby anchoring the security in the hands of other external companies.}. From these promising directions of research, it may seem that quantum key distribution systems will become nearly perfect in the future, in the sense that all imperfections are either eliminated, or accounted for by additional privacy amplification as quantified by security proofs.

In other words, the eavesdropper Eve in QKD seems to have a sad destiny. She initially had two tools in her suitcase: attacking perfect QKD systems with optimal quantum attacks, and quantum hacking attacks exploiting imperfections. The security proofs eliminated the first tool, while the recent developments in implementations and practical security proofs are about to eliminate the second. However, in this Letter we demonstrate a third tool in her suitcase. Eve may intentionally damage the system, to actively engineer exploitable imperfections. In this way, even an initially perfect setup can become totally insecure, without raising any alarms. This clearly demonstrates the fact that it is not sufficient to have well-characterized components and systems. Eve may totally change their behavior at some later point. The results ultimately question if communication security is physically attainable at all, in principle. 

Changes in characteristics of most optical components inside a QKD system can lead to loopholes being created. QKD schemes rely on known characteristics of, for example, attenuators, beamsplitters, modulators, polarization control components, spatial and spectral filters, optical connectors, lenses, mirrors, light sources and detectors. For a proof-of-principle demonstration of the new class of attacks, we needed to pick a target component and a target type of QKD system. A natural choice was avalanche photodiode (APD) in a free-space system, for the following reasons. A high-power laser beam is experimentally easier to apply through free-space optics. The APD absorbs most of the incoming light in a small area, which makes it likely to suffer damage at lower power than other optical components. We decided to investigate a widely used Si APD (PerkinElmer C30902SH), employed in single-photon detectors in several QKD experiments \cite{kurtsiefer2002,schmitt-manderbach2007,peloso2009,hughes2002,poppe2004.OptExpress-12-3865,kurochkin2005.TechPhys-50-727,duligall2006.NewJPhys-8-249}. For this component, we have demonstrated permanent laser damage useful for eavesdropping, as detailed below.

\begin{figure}
\includegraphics[width=\columnwidth]{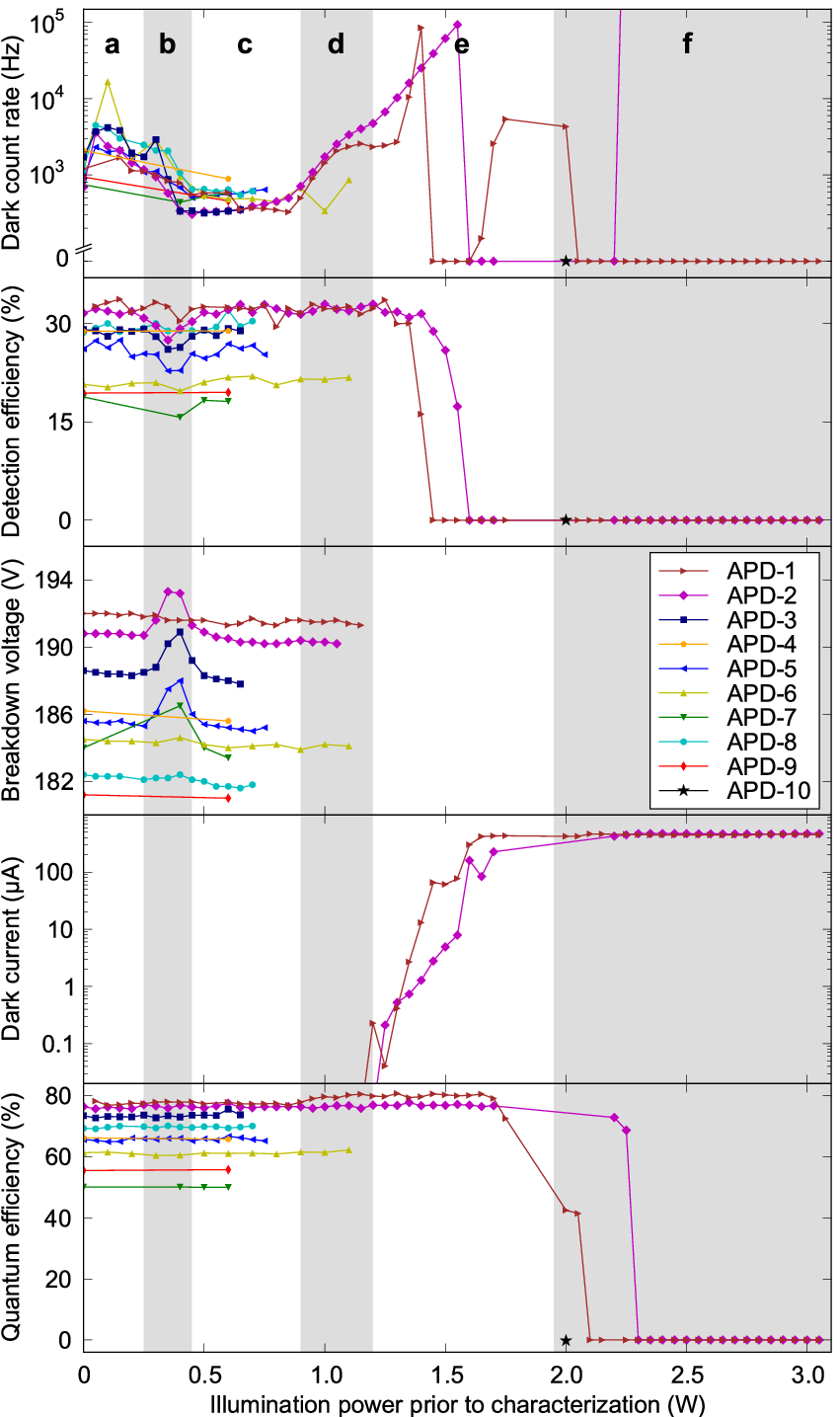}
\caption{\label{fig:dmg-results} Results of applying high-power illumination to ten APD samples. The data points show APD and detector parameters measured after each successive application of illumination of increasing peak power. The leftmost point on each trace is the initial value of parameter prior to illumination.}
\end{figure}

Initial tests showed that useful laser damage could be achieved. For thorough characterization of effects we subsequently built an automated setup (Fig.~\ref{fig:setup}) that applied damaging light in small increments, and fully characterized the APD in between the exposures \cite{ghazali2011.IIUMEngJ-12-no5-97}. The setup tests a stand-alone APD, however the results are applicable to a complete QKD system as discussed through this Letter. High-power continuous-wave (c.w.)\ illumination is produced by electrically controlled $807\,\nano\meter$ laser diode, pigtailed with multimode (MM) fiber of $200\,\micro\meter$ core diameter. The beam exiting the MM fiber is collimated, passes through a 50:50 non-polarizing beamsplitter (diverting half of the power into a power meter), a mechanical shutter, and is focused at the APD in $50\,\micro\meter$ full-width at half-maximum (FWHM) diameter spot. In an actual attack, the wavelength of damaging laser would have to be close to that of signal photons, because all free-space QKD systems employ a narrowband interference filter at the entrance to cut background light in the channel. Many of these systems operate in $770$--$850\,\nano\meter$ wavelength range \cite{kurtsiefer2002,schmitt-manderbach2007,peloso2009,hughes2002,poppe2004.OptExpress-12-3865,kurochkin2005.TechPhys-50-727}, not far from the damaging laser wavelength in our test.

In addition to the high-power laser, our setup has two single-mode (SM) fiber pigtailed lasers and a variable attenuator. These provide calibrated pulsed and c.w.\ illumination for characterizing the APD. The APD is connected into a standard passively-quenched single-photon detector scheme, and thermoelectrically chilled to $-25\,\celsius$ \cite{cova1996,kim2011.RevSciInstrum-82-093110}. The detector output is connected to a counter and time interval analyzer. Bias is applied to the APD from a programmable voltage source (HV), allowing measurement of I--V curves and several electrical and optical characteristics \cite{ghazali2011.IIUMEngJ-12-no5-97}. We measured detector dark count rate and photon detection efficiency with APD biased $15\,\volt$ above its initial (undamaged) breakdown voltage value $V_\text{br orig.}$, APD breakdown voltage $V_\text{br}$, dark current when biased in linear amplification mode $5\,\volt$ below $V_\text{br orig.}$, and photoconversion quantum efficiency when biased at $0\,\volt$.

Most of our tests proceeded by applying a cycle of c.w.\ illumination for $60\,\second$, then characterizing the APD. The power level was increased in small increments between the cycles. The software paused the experiment and alerted the operator if any characterized parameter deviated significantly from its initial value. Then the operator would either continue the test to higher powers and further destruction of the sample, or terminate the test to check for sample's long-term stability. Results of the tests are plotted in Fig.~\ref{fig:dmg-results}.

We tested 10 samples of PerkinElmer C30902SH APD in total, numbered APD-1 through APD-10 in this Letter. The samples came from different production batches manufactured in 2009--2010. The changes observed in all samples after high-power illumination were generally consistent between the samples and permanent, unless noted otherwise. As we applied illumination of increasing power, we observed seven distinct effects denoted by vertical bands \textbf{a--f} in Fig.~\ref{fig:dmg-results} and explained below.

\textbf{a.}~After illumination of less than $0.25\,\watt$ power, dark count rate of the APDs rose by several times. This is the only non-permanent effect, dissipating after the APD is left in darkness for several hours. (Also the only known, noted in the APD data sheet.)

\textbf{b.}~In $0.3$ to $0.45\,\watt$ range, 4 out of 8 APDs tested in this range exhibited rise of their $V_\text{br}$ by $2.3$--$2.5\,\volt$. This was accompanied by a reduction in their photon detection efficiency by a factor of $0.83$--$0.90$. Hypothesized mechanism of the efficiency reduction is that while the APDs are biased at a constant voltage in the detector, the rise of $V_\text{br}$ lowered overvoltage (the difference between the bias voltage and $V_\text{br}$), leading to lower detection efficiency \cite{dautet1993,kim2011.RevSciInstrum-82-093110}. When attacking a complete Bob, Eve could thus reduce sensitivity of a selected APD. This is because individual APDs are addressable by varying polarization or other parameters of the damaging light at Bob's input. This would create a permanent efficiency mismatch between Bob's detectors \cite{makarov2006}. This efficiency mismatch can potentially increase Eve's knowledge of the key, if Bob does not recharacterize his detectors or accounts for such imperfections in the postprocessing procedure.

\textbf{c.}~In $0.5$ to $0.8\,\watt$ range, all APD parameters returned to normal, with the exception of the dark count rate that remarkably fell 1.7--5.4 times comparing to the original dark count rate measured before starting the treatment. The dark count rate reduction was observed in all 8 samples tested in this power range, and the change has been verified to be permanent. This is to our knowledge the first demonstration that Eve can \emph{improve} legitimate user's equipment. The default treatment of all errors in QKD is that they resulted from eavesdropping, regardless of the actual error source. Detector dark counts therefore limit the maximum transmission distance of a given system, raising the quantum bit error rate (QBER) beyond the secure limit as the photon transmission probability drops. With some extra assumptions or complications in the detection setup, it is possible to improve QKD performance beyond this limit \cite{qi2007.PhysRevA-75-052304,scarani2009}. Similarly, it is tempting to simply subtract a calibrated dark count rate from the QBER. Our result clearly shows that this can be dangerous; all errors in the raw key must be treated as caused by eavesdropping \cite{scarani2009}.

\textbf{d.}~In $0.9$ to $1.2\,\watt$ range, the dark count rate permanently rose to large values.

\begin{figure}
\includegraphics{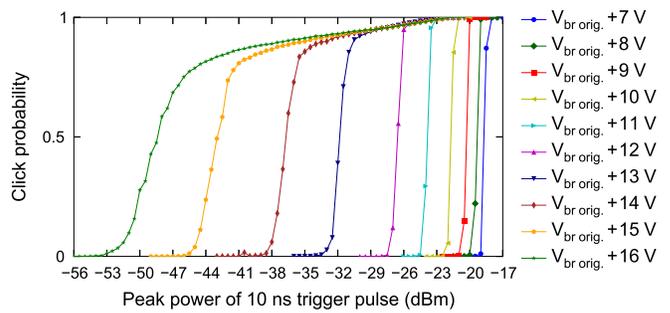}
\caption{\label{fig:control-characteristics} Detector control characteristics of a permanently blinded APD-1, at different overvoltage values. Note that trigger pulse power needs to change less than $3\,\deci\bel$ (i.e., less than 2 times) for a change of click probability from $0$ to $>$$0.5$, at typical operating overvoltages of the APD in $10$--$15\,\volt$ range. This would allow a perfect or near-perfect faked-state attack on QKD system \cite{makarov2008a-v1,lydersen2010a,lydersen2011}. Note that perfect deterministic 0-or-1 click probability control, as evident at overvoltages $\leq 11\,\volt$, is not required for a successful attack. Even probabilistic control at larger overvoltages should suffice to break security in most if not all practical settings \cite{lydersen2011b}.}
\end{figure}

\textbf{e.}~In $1.2$ to $1.7\,\watt$ range, the APDs developed large dark current. This led to blinding of the passively-quenched detector, dropping the photon detection efficiency and dark count rate to zero in both samples tested in this range. The blinding mechanism is that excessive current drawn by the APD from the bias circuit (in our case from $400\,\kilo\ohm$ ballast resistor) leads to the voltage supplied by the circuit dropping below $V_\text{br}$, as previously demonstrated by weaker c.w.\ illumination \cite{makarov2009}. The difference here is that the laser damage blinding is permanent and does not require continuing illumination. Under the blinded condition the detector remains photosensitive to moderately bright light and is either perfectly controllable or well-controllable (depending on overvoltage operating setpoint) by $10\,\nano\second$ wide light pulses, see Fig.~\ref{fig:control-characteristics}. This renders it insecure for QKD applications \cite{lydersen2010a,gerhardt2011,lydersen2011}.

\textbf{f.}~At $\geq 2\,\watt$, catastrophic structural damage took place. We tested 3 samples to this power range. In one of them (APD-10, single experimental point at $2\,\watt$ in Fig.~\ref{fig:dmg-results}) the bonding wires melted off, leaving the device an open circuit. The other two reduced then completely lost all photosensitivity, with the device becoming a resistor in $10$--$100\,\kilo\ohm$ range. If this APD were employed for a watchdog power meter as in one countermeasure proposal \cite{lydersen2010a}, the countermeasure would be defeated.

\begin{figure}
\includegraphics{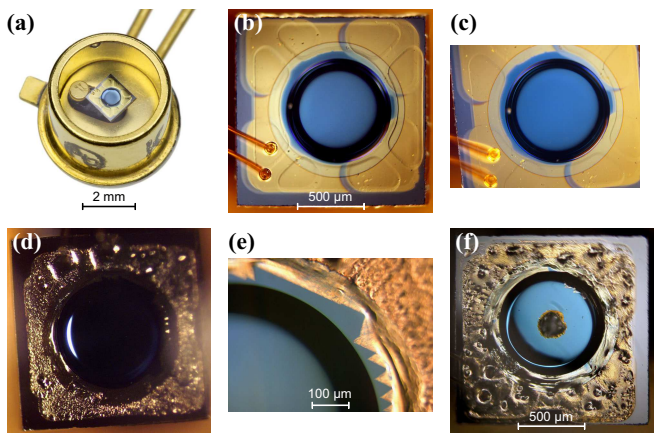}
\caption{\label{fig:apd-images} Microscope images of APDs at various stages of damage. (a) APD package with Si chip behind a glass window; the other images show chip close-ups. (b) Untreated APD-3. (c) APD-3 after $0.65\,\watt$ illumination, which has reduced the dark count rate and produced no visible damage. (d)--(e) APD-1 after $2\,\watt$ illumination, showing re-melted gold electrode and gold flowing into clear area along Si crystal lattice planes; this sample has large dark current but unchanged quantum efficiency. (f) APD-2 after $3\,\watt$ illumination with a hole blown in the middle through the entire thickness of the silicon chip; it has zero photosensitivity, resistor-like state. Damaging illumination in all cases was applied for $60\,\second$. Images (b)--(e) were taken with bright-field illumination, (d) chip surface intentionally tilted, (f) dark-field illumination.}
\end{figure}

Later stages of damage result in visible changes to the APD chip (Fig.~\ref{fig:apd-images}). The first visible change is disfiguring of the gold electrode, possibly resulting from Si--Au alloy formation at $>370\,\celsius$ \cite{haitz1965}. In the last stage of damage, the laser beam always produces a hole in Si chip.

The permanent reduction of dark count rate is an interesting effect. We tested most of our samples illuminated with $50\,\micro\meter$-focused, $60\,\second$ square pulses of successively increasing power levels, and kept the detector high-voltage source at $V_\text{br orig.}+15\,\volt$ through the test. However we have also tested with a single $60\,\second$ square pulse applied to a fresh sample (APD-7); with illumination slowly linearly ramped up in $900\,\second$, kept constant for $60\,\second$ then linearly ramped down to zero in $900\,\second$ (APD-4); with illumination defocused such that the spot became larger than the APD photosensitive area (APD-6); finally, with the high-voltage source switched off for the duration of laser treatment (APD-8). In all cases we observed permanent reduction of dark count rate. It appears that the main cause of it is heating the APD chip to a certain peak temperature. A similar effect has previously been observed and attributed to localized annealing when APD junction was heated by electrical current \cite{haitz1965}.

%Discussion

The results of testing this component clearly support that Eve may, in general, alter the system characteristics by altering characteristics of its optical components. Then the system no longer complies with the security proof. Then, even with a sufficiently general security proof, and with a QKD implementation that is pre-characterized to comply with the security proof, security cannot be guaranteed. The countermeasure can be to characterize the system more frequently to ensure the validity of the characterization. One could imagine doing this whenever an unusual event was detected, as the bright power of the damaging laser surely has a temporary signature on the system. Meanwhile, it is difficult to exhaustively list all events that should trigger a recharacterization. Eve could for instance wait for a power outage, and perform the damage when the system is unpowered.

It is therefore advisable to monitor the characteristics of the system directly during QKD, or at least such that the characteristics are bounded during QKD with a sufficiently high probability. Thus the security proofs should minimize the number of necessary characteristics about the system. One example is the Bennett-Brassard-Mermin 1992 (BBM92) scheme where the source of entangled photons does not need to be characterized \cite{bennett1992a}. Another example is the proof for measurement-device-independent QKD systems \cite{lo2012} that has no necessary characterized parameters for the Bell-state analyzer including the detectors. Yet another example is the security proof in Ref.\ \cite{maroy2010}, where the secure key generation rate is only dependent on one imperfection parameter at Bob's side, namely the minimum detection efficiency of a nonvacuum state incident to Bob.

On the implementation side, it turns slightly into a cat-and-mouse game, where Alice and Bob must ensure that the in-field characterization during QKD is reliable and untampered by Eve. Optical power limiters is a well-studied technology that may be applied against tampering at the entrances of Alice and Bob \cite{tutt1993.ProgQuantElectr-17-299}, and using a watchdog power meter has been proposed \cite{makarov2005,lydersen2010a}. However our results clearly show that Eve might tamper with these countermeasures. In a more narrow example, detectors can be tested for single-photon sensitivity at random times to bound the minimum detection efficiency \cite{lydersen2011a}. Again, to do this in-field is not trivial, and the security then again relies on the pre-characterization of the single-photon source and path used for testing. A reliable in-field scheme to characterize crucial equipment parameters during operation can be a future study.

Finally, our study shows the practical challenge of physically securing a QKD system from all side-channels. This is one of the most fundamental assumptions in most security proofs (even in the device-independent security proofs \cite{acin2007}), and possibly the hardest to fully characterize. For example, one can envision a situation where Eve damages the detectors or other crucial components, not by using the fiber, but rather by focusing high-power X-ray radiation onto the components from outside of the system. Another, probably future way to gain access could be a nanorobot burrowing through the fiber core.

\begin{acknowledgments}
We thank E.~Anisimova for valuable assistance during the experiments; C.~Kurtsiefer, Y.-S.~Kim and Q.~Liu for sharing electronics and mechanical design used in parts of the experimental setup. This work was supported by the Research Council of Norway (grant no.~180439/V30), University Graduate Center in Kjeller, and Industry Canada.
\end{acknowledgments}

\end{document}